\documentclass{jpp}
\usepackage{subeqn}
\usepackage{epsfig}

\let\realverbatim\verbatim
\let\realendverbatim\endverbatim
\renewenvironment{verbatim}
  {\par\addvspace{6pt plus 1pt minus 2pt}\realverbatim}
  {\realendverbatim\addvspace{6pt plus 1pt minus 2pt}}

\ifprodtf \else
  \checkfont{eurm10}
  \iffontfound
    \IfFileExists{upmath.sty}
      {\typeout{^^JFound AMS Euler Roman fonts on the system,
                   using the 'upmath' package.^^J}%
       \usepackage{upmath}}
      {\typeout{^^JFound AMS Euler Roman fonts on the system, but you
                   don't seem to have the}%
       \typeout{'upmath' package installed. JPP.cls can take advantage
                 of these fonts,^^Jif you use 'upmath' package.^^J}%
       \providecommand\umu{\umu}%
      }
  \else
    \providecommand\umu{\mu}%
  \fi
\fi

\ifprodtf \else
  \checkfont{msam10}
  \iffontfound
    \IfFileExists{amssymb.sty}
      {\typeout{^^JFound AMS Symbol fonts on the system, using the
                'amssymb' package.^^J}%
       \usepackage{amssymb}%

      }{}
  \else
  \fi
\fi

\ifprodtf \else
  \IfFileExists{amsbsy.sty}
    {\typeout{^^JFound the 'amsbsy' package on the system, using it.^^J}%
     \usepackage{amsbsy}}
    {}
\fi

\newcommand{\refeqb}[1]{Equation (\ref{#1})}

\newcommand{\refeqm}[1]{Eq. (\ref{#1})}

\newcommand{\Fig}[1]{Fig. (\ref{#1})}
\newcommand{\Figb}[1]{Figure (\ref{#1})}

\newcommand{\eqns}[2]{Eqs. (\ref{#1}){ \& (\ref{#2}})}
\newcommand{\beqns}[2]{Equations (\ref{#1}){ \& (\ref{#2}})}

\newcommand{\figs}[2]{Figs. (\ref{#1}){ \& (\ref{#2}})}

\newcommand{\be}{\begin{equation}}
\newcommand{\ee}{\end{equation}}

\newcommand{\eqa}{\begin{eqnarray*}}
\newcommand{\eqe}{\end{eqnarray*}}
\newcommand{\eqnu}{\begin{eqnarray}}
\newcommand{\eqne}{\end{eqnarray}}

\newcommand{\eqs}[2]{Eqs. (\ref{#1}) \& (\ref{#2})} 

\newcommand{\eq}[1]{Eq. (\ref{#1})} 

\newcommand{\eeq}{\end{eqnarray}}  

\newdefinition{definition}[theorem]{Definition}

\title[Journal of Plasma Physics]{Theory and Simulations of Whistler Wave Propagation}

\author[Dastgeer Shaikh]
{D\ls A\ls S\ls T\ls G\ls E\ls 
E\ls R\ns S\ls H\ls A\ls I\ls K\ls H\ls }
\affiliation{Institute of Geophysics and Planetary Physics (IGPP),\\
University of California, Riverside, CA 92521. USA.\\
{\tt Email:dastgeer@ucr.edu}}
\date{15 Feb 2008, Revised 25 Feb 2008}
\pubyear{2008} 
\volume{00}
\part{0}
\pagerange{\pageref{firstpage}--\pageref{lastpage}}
\doi{S0963548301004989}

\begin{document}

\label{firstpage}
\maketitle

\begin{abstract}
A linear theory of whistler wave is developed wihtin the paradigm of a
two dimensional incompressible electron magnetohydrodynamics
model. Exact analytic wave solutions are obtained for a small
amplitude whistler wave that exhibit magnetic field topological
structures consistent with the observations and our simulations in
linear regime.  In agreement with experiment, we find that the
parallel group velocity of the wave is large compared to its
perpendicular counterpart. Numerical simulations of collisional
interactions demonstrate that the wave magnetic field either coalesces
or repels depending upon the polarity of the associated current. In
the nonlinear regime, our simulations demonstrate that the evolution
of wave magnetic field is governed essentially by the nonlinear Hall
force.
\end{abstract}


\section{Introduction}
Whistler waves are widely observed in many space and laboratory plasma
phenomena. For instance, they are believed to be generated in the
Earth's ionospheric region by lightning discharges and proceed in the
direction of Earth's dipole magnetic field [\cite{helli}]. They have
also been recently detected in the Earth's radiation belt by the
STEREO S/WAVES instrument [\cite{cattell}].  Moreover, there are
observations of Venus' ionosphere that reveal strong, circularly
polarized, electromagnetic waves with frequencies near 100Hz. The
waves appear as bursts of radiation lasting 0.25 to 0.5s, and have the
expected properties of whistler-mode signals generated by lightning
discharges in Venus' clouds [\cite{Russell}]. These waves are also
reported near the Earth's magnetopause [\cite{Stenberg}] and the
Cluster spacecraft encountered them during the process of magnetic
reconnection in the Earth's magnetotail region [\cite{wei}].  Upstream
of collisionless shock, whistler waves are found to play a crucial
role in heating the plasma ions [\cite{Scholer}]. Their excitation and
propagation are not only limited to the Earth's nearby ionosphere, but
they are also found to be excited near the ionosphere of other planets
such as in the radiation belts of Jupiter and Saturn
[\cite{Bespalov}].  Whistler waves are believed to be a promising
candidate in transporting fields and currents in plasma opening switch
(POS) devices [\cite{mason}], which operate on fast electron time
scales. Similarly, these waves have been known to drive the phenomenon
of magnetic field line reconnection [\cite{bul}] in astrophysical
plasmas [\cite{zhou}].  Whistler waves have also been investigated in
several laboratory experiments
[\cite{sten1,sten2,sten3,sten4,sten5,sten6}], where they have been
found to exhibit a variety of interesting features such as anisotropic
propagation of phase front, strong dispersion characteristics,
interaction with plasma particles etc.  A few experimental features
have also been confirmed by  recent three dimensional simulations
[\cite{eliasson}], where it has been reported that the polarity and
the amplitude of the toroidal magnetic field, in agreement with the
laboratory experiments, determine the propagation direction and speed
of the whistlers.  These are only a few examples amongst a large body
of work devoted to the study of whistler waves.  Despite the large
amount of efforts gone into understanding the existence and
propagation of whistlers, their linear dynamics is still debated,
specially in the context of complex nonlinear processes. For example,
the role of whistler waves in high frequency turbulence and
anisotropic spectral cascades has been debated recently
[\cite{dastgeer1,dastgeer2,dastgeer3,dastgeer4,dastgeer5}].

Motivated by these issues, we in this paper develop a fully
self-consistent linear analytic theory of whistler wave propagation.
The electromagnetic whistler waves in a plasma are excited essentially
by means of collective electron oscillations in presence of an
external or self-consistent large scale magnetic field. Their
characteristics frequency thus lies between ion- and
electron-gyrofrequencies i.e., $\omega_{ci}\ll\omega\ll\omega_{ce}$,
where $\omega_{ci}$ and $\omega_{ce}$ are ion- and
electron-gyrofrequencies respectively. In such a frequency regime,
usual magnetohydrodynamic theory becomes invalid and the plasma
processes need to be described by electron magnetohydrodynamics (EMHD)
theory [\cite{kingsep}]. The linear properties of EMHD are mainly
governed by whistler waves, which constitute fundamental oscillatory
modes of this model.

In this paper, we develop a linear analytical theory and simulation to
investigate the linear propagation of a whistler wave packet and
demonstrate that such waves lead to an exact solution of EMHD
model. For the sake of simplicity, we restrict ourselves to a two
dimensional plane where variations in the wave magnetic field are
confined to the $x$ and $y$ directions only.  The analytic results of
our theory show a qualitative/quantitative agreement with the
experimental observations and simulations.  Such a comparison further
enables us explain several experimentally observed features of the
propagating whistler wave packets.  We also carry out an investigation
of collision of two propagating whistler wave packets to explore their
mutual interaction in the context of ionospheric processes.  The
remainder of this paper is organized as follows. In Sec 2, governing
equations of whistler waves and dispersive properties are described.
In Sec 3, we describe the evolution of wave amplitude based on Fourier
analytic method and numerical simulations.  Section 4 describes the
propagation studies of whistlers that show a remarkable agreement with
the simulation in the linear regime. The linear simulations are
discussed in Sec 5. Section 6 contains results of mutual interaction
of two whistler modes, whereas Sec 7 describes nonlinear evolution of
the whistlers. Finally conclusions are discussed in Sec 8.

\section{Dynamical Equations of Whistler Waves}
Whistler wave dynamics is governed essentially by the equations of
EMHD.  The EMHD phenomena occur typically on rapid electron time
scales [\cite{kingsep}], while ions do not participate in the
dynamics. Thus, the basic frequency scales involved are
$\omega_{ci} \ll \omega \ll \omega_{ce}$ (where $\omega_{ci},
\omega_{ce}$ are respectively ions and electrons gyro-frequencies, and
$\omega$ is the characteristic frequency) and the length scales are
$c/\omega_{pi} < \ell < c/\omega_{pe}$, where $\omega_{pi},
\omega_{pe}$ are the plasma ion and electron frequencies. 
Currents carried by the electron fluid are important. Since stationary ions 
merely provide a neutralizing background to a quasi-neutral EMHD plasma, their
momentum can be ignored in the high frequency regime.
The momentum
equation of electron is
\be
\label{elec}
m_e n \left( \frac{\partial {\bf V}_e}{\partial t} + {\bf V}_e \cdot \nabla {\bf V}_e \right)
=-en {\bf E} -  \frac{ne}{c} {\bf V}_e \times {\bf B} - \nabla P
-\mu m_e n {\bf V}_e, 
\ee
\be
{\bf E} = -\nabla \phi - \frac{1}{c} \frac{\partial {\bf A}}{\partial t},
\ee
\be
\label{ampere}
\nabla \times {\bf B} = \frac{4\pi}{c} {\bf J} +  
\frac{1}{c} \frac{\partial {\bf E}}{\partial t},
\ee
\be
\frac{\partial n}{\partial t} + \nabla \cdot (n {\bf V}_e) = 0.
\ee
The remaining equations are ${\bf B} = \nabla \times {\bf A}, {\bf J}
= -en{\bf V}_e, \nabla \cdot {\bf B} =0$. Here $m_e, n, {\bf V}_e$ are
the electron mass, density and fluid velocity respectively. ${\bf E},
{\bf B}$ respectively represent electric and magnetic fields and $\phi
, {\bf A}$ are electrostatic and electromagnetic potentials. The
remaining variables and constants are, the pressure $P$, the
collisional dissipation $\mu$, the current due to electrons flow ${\bf
J}$, and the velocity of light $c$.  The displacement current in
Ampere's law \eq{ampere} is ignored, and the density is considered as
constant throughout the analysis.  The electron continuity equation
can therefore be represented by a divergence-less electron fluid
velocity $\nabla \cdot {\bf V}_e = 0$. The electron fluid velocity can
then be associated with the rotational magnetic field through
\[
{\bf V}_e = - \frac{c}{4\pi n e}\nabla \times{\bf B}.
\]
On taking the  curl of \eq{elec} and, after slight rearrangement
of the terms, we obtain a generalized electron momentum equation
in the following form.
\be
\label{PP}
\frac{\partial {\bf P}}{\partial t} - {\bf V}_e  \times ( \nabla \times {\bf P})
+\nabla \xi = - \mu m_e {\bf V}_e
\ee
where
\[
{\bf P} = m_e{\bf V}_e - \frac{e{\bf A}}{c} ~~~~~~~{\rm and}~~~
\xi = \frac{1}{2} m_e {\bf V}_e\cdot {\bf V}_e + \frac{P}{n} - e\phi.
\]
The curl of \eq{PP} leads to 
a three-dimensional equation of EMHD describing the evolution of
the whistler wave magnetic field, 
\be
\label{emhd3}
\frac{\partial }{\partial t}({\bf B}-d_e^2 \nabla^2 {\bf B} ) + {\bf
V}_e\cdot \nabla ({\bf B}-d_e^2 \nabla^2 {\bf B} )- ({\bf B}-d_e^2
\nabla^2 {\bf B} ) \cdot \nabla{\bf V}_e= \mu d_e^2 \nabla^2 {\bf B}.
\ee 
where $d_e=c/\omega_{pe}$, electron skin depth, is an intrinsic
length-scale in EMHD.  The three-dimensional equations of EMHD can be
transformed into two dimensions by regarding variation in the
$\hat{z}$-direction as ignorable i.e.  $\partial/\partial z=0$, and
separating the total magnetic field ${\bf B}$ into two scalar
variables, such that ${\bf B} = \hat{z} \times \nabla \psi +
b\hat{z}$.  Here $\psi$ and $b$ respectively present perpendicular and
parallel components of the wave magnetic field. The corresponding
equations of these components can be written in a normalized form as follows,
\be
\label{psi}
 \frac{\partial}{\partial t}(\psi - d_e^2\nabla^2 \psi) +
{\hat{z}\times{\bf \nabla}b} \cdot \nabla (\psi - d_e^2\nabla^2 \psi) 
- B_0\frac{\partial}{\partial y}b
=0,
\ee 
\be 
\label{b}
\frac{\partial}{\partial t} (b - d_e^2\nabla^2 b) -
d_e^2{\hat{z}\times{\bf \nabla}b} \cdot \nabla \nabla^2 b +
{\hat{z}\times{\bf \nabla}\psi} \cdot \nabla \nabla^2 \psi 
+ B_0 \frac{\partial}{\partial y}\nabla^2 \psi 
=0 .  
\ee
The length and time scales are normalized respectively by $d_e$ and
$\omega_{ce}$, whereas magnetic field is normalized by a typical mean
$B_0$. The linearization of \eqs{psi}{b} about a constant magnetic
field $B_0$ yields the dispersion relation for the whistlers, the
normal mode of oscillation in the EMHD frequency regime, and is given
by
\[
\omega_k = \omega_{c_0}\frac{d_e^2 k_yk}{1+d_e^2k^2},
\]
where $ \omega_{c_0}=eB_0/mc$ and $k^2=k_x^2+k_y^2$.  From the set of
 the EMHD \eqs{psi}{b}, there exists an intrinsic length scale $d_e$,
 the electron inertial skin depth, which divides the entire spectrum
 into two regions; namely short scale ($kd_e>1$) and long scale
 ($kd_e<1$) regimes.  In the regime $kd_e<1$, the linear frequency of
 whistlers is $\omega_k \sim k_y k$ and the waves are dispersive.
 Conversely, dispersion is weak in the other regime $kd_e>1$ since
 $\omega_k \sim k_y/ k$ and hence the whistler wave packets interact
 more like the eddies of hydrodynamical fluids.

\section{Evolution of Whistler Wave Amplitude}
The evolution of the amplitude associated with a linear whistler wave
can be studied from \eqs{psi}{b} after dropping the nonlinear terms,
as described in the following.
\begin{equation}
\label{psi-eqn}
\frac{\partial}{\partial t}(\psi - \nabla^2 \psi)
 - B_0\frac{\partial b}{\partial y}
 = \mu\nabla^2 \psi
\end{equation}
\begin{equation}
\label{b-eqn}
\frac{\partial}{\partial t}(b  - \nabla^2 b)
+ B_0 \frac{\partial}{\partial y}\nabla^2 \psi = \mu \nabla^2 b.
\end{equation}
The rhs of the above equations now contains a damping term.
We adopt following definition to Fourier transform the evolution
variables of \eqs{psi-eqn}{b-eqn},
\be
\label{ft}
G(x,y,t) =\frac{1}{2\pi} \int_{-\infty}^{+\infty}\int_{-\infty}^{+\infty}
 G(k_x,k_y,t) \exp(ik_x x+ik_y y) dk_x dk_y 
\ee
and the corresponding inverse Fourier transform is
\be
\label{ift}
G(k_x,k_y,t) =\frac{1}{2\pi}
\int_{-\infty}^{+\infty}\int_{-\infty}^{+\infty} G(x,y,t) \exp(-ik_x
x-ik_y y) dx dy.
\ee
We then write \eqs{psi-eqn}{b-eqn} as,
\be
\label{mateqn}
\frac{\partial}{\partial t} \left(\begin{array}{c}
b_k \\[0.5cm] \psi_k \end{array} \right)  =
\left(\begin{array}{c} \frac{-\mu k^2}{1+k^2}\hspace{6.0mm} 
\frac{ik_y B_0k^2}{1+k^2}
\\[0.7cm] \frac{ik_y B_0}{1+k^2} \hspace{6.0mm} \frac{-\mu k^2}{1+k^2} \end{array}
\right)  \left(\begin{array}{c}
b_k \\[0.5cm] \psi_k \end{array} \right) 
\ee
where $~b_k \sim b(k_x,k_y,t)$ and $\psi_k \sim \psi(k_x,k_y,t)$. The
solutions of Eq. (\ref{mateqn}) can be written as;
\be
\label{bk}
b_k(t) = \frac{1}{2}\left[b_k(0)+k\psi_k(0)\right] \exp(\lambda_+ t) 
       + \frac{1}{2}\left[b_k(0)-k\psi_k(0)\right]\exp(\lambda_- t)  
\ee
\be
\label{psik}
\psi_k(t) = \frac{1}{2k}\left[b_k(0)+k\psi_k(0)\right] \exp(\lambda_+ t) 
       - \frac{1}{2k}\left[b_k(0)-k\psi_k(0)\right]\exp(\lambda_- t)  
\ee
where $b_k(0)$ and $\psi_k(0)$ denote the Fourier transformed initial
conditions of the variables $b(x,y,t=0)$ and $\psi(x,y,t=0)$
respectively, and $\lambda$ is an eigen value of the squared matrix of
 Eq. (\ref{mateqn}) which is given as below;
\be
\label{eigen}
\lambda_{\pm} = \frac{-\mu k^2}{1+k^2}\pm\frac{ik_y k B_0}{1+k^2}.
\ee
Here also, $\pm$ sign corresponds to the wave moving in the forward
and backward direction. To seek an analytic solution
of \eqs{bk}{psik}, we initialize the wave magnetic field perturbations
by means of a current carrying antenna similar to that of Stenzil's et
al [\cite{sten5}] experiment such that there exists only in plane
component of the wave magnetic field, whereas the component along the
ambient magnetic field is zero initially.  Accordingly, we choose
the following kind of initial perturbation to study the behaviour of wave
amplitude
\eqa
\psi(x,y,t=0)&=&\Psi_0\cos(ax)\exp\left(-\frac{y^2}{\sigma_y^2}\right) \\ \nonumber
 b(x,y,t=0)   &=& 0 
\eqe
Fourier transformation yields,
\eqnu
\label{ini1}
\psi_k(0) &=& \frac{\Psi_0 \sigma_y}{2}
\sqrt{\pi}[\delta (k_x+a) +
\delta (k_x-a)] \exp \left(-\frac{\sigma_y^2 k_y^2}{4} \right) \\ \nonumber
b_k(0) &=& 0
\eqne
We would like to concentrate on the case in which a wave packet is
propagating in the forward direction only, and hence choose the eigen
value that corresponds to $\lambda_+$ in the Eq. (\ref{eigen}). Under
such conditions, the poloidal component of wave magnetic field i.e.,
\refeqm{psik} assumes the form as;
\[ \psi_k(t) = \frac{1}{2} \psi_k(0)  \exp(\lambda_+ t) \]
substituting \refeqm{ini1} into above eqn and then inverse Fourier
transformation gives
\be
\label{psi-int}
 \psi(x,y,t) = \frac{\Psi_0 \sigma_y \sqrt{\pi}}{8\pi}
\int_{-\infty}^{+\infty}[\delta (k_x+a) +\delta (k_x-a)]\exp(ik_x x)\zeta(k_x)
dk_x
\ee
where,
\be
\label{spa}
\zeta(k_x) = \int_{-\infty}^{+\infty}
\exp \left(-\frac{\sigma_y^2 k_y^2}{4} \right)
\exp \left( i B_0 k k_y t \right)\exp(ik_y y) dk_y,
\ee
The expression of $\psi$ contains another integral given by
\refeqm{spa}. We therefore first solve \refeqm{spa} and substitute the
solution into
\refeqm{psi-int}. In order to do so, we put \refeqm{spa} into the 
following form
\be
\label{st}
\zeta(k_x) = \int_{-\infty}^{+\infty}F(k) \exp \left[ i \chi(k) t \right] dk_y
\ee
where $k^2 =(k_x^2+k_y^2)$ and
\[ F(k) = \exp \left(-\frac{\sigma_y^2 k_y^2}{4} \right) \]
\[ \chi(k) = \omega(k) +k_y \frac{y}{t} \]
\[  \omega(k) = B_0 k_y k \]
We employ the method of stationary phases [\cite{arfken,witham}] to analyze
\refeqm{st}.  This method indicates that in the time asymptotic limit
(i.e., at large times), the dominant contribution in the integral will
essentially come from the maximum of the exponential part. This
further suggests that the function $\chi(k)$ in \refeqm{st}, must take
on a real and positive maximum value. Away from this maximum, the
integrand will have negligibly small contribution. The condition of
maximization of $\chi(k)$ further implies that
\be 
\label{extrem}
\frac{\partial}{\partial k_y} \chi(k) = 0.
\ee
The vanishing of first derivative reveals that we do have stationary
points, which may be either maximum or minimum. \refeqb{extrem} gives
a bi-quadratic algebraic equation in $k_y$, whose roots are basically
functions of $k_x$.
\[ k_y(k_x) = \pm \left \{\frac{y^2}{8B_0^2 t^2}-\frac{k_x^2}{2} \pm 
\frac{y}{2B_0 t} \sqrt{ \frac{y^2}{16 B_0^2 t^2}+\frac{k_x^2}{2}} 
\right \}^{1/2} \]
The above equation admits four roots.  We thus have four stationary
points corresponding to these four roots.  We take into account only
that point, which gives a maximum contribution in the \refeqm{st}. We
denote this point as $k_0 = k^{\rm max}_y(k_x)$. Taylor expanding the
function $\chi(k)$ around this stationary point ($k_0$), we may write
\[  \chi(k) = \chi(k_0) + \underbrace{(k_y-k_0) \chi'(k_0)}_0 + 
\frac{1}{2!}(k_y-k_0)^2 \chi''(k_0). \]
Note the second term in the expansion is zero because of the condition
of extremum imposed by \refeqm{extrem}. Substituting $\chi(k)$ in
\refeqm{st}, we get
\[ \zeta(k_x) = F(k_0) \exp \left[i\chi(k_0)t\right]\int_{-\infty}^{+\infty} 
\exp \left[ i \frac{1}{2!}(k_y-k_0)^2 \chi''(k_0)  t \right] dk_y. \]
\refeqb{psi-int} with the help of above expression give,
\eqnu
\psi(x,y,t) = \frac{\Psi_0 \sigma_y \sqrt{\pi}}{8\pi}
\int_{-\infty}^{+\infty}[\delta (k_x+a) +\delta (k_x-a)]\exp(ik_x x)\nonumber\\  
\times F(k_0) \exp \left[i\chi(k_0)t\right]
\sqrt{\frac{2\pi}{t\chi^{\prime \prime}(k_0)}} dk_x
\eqne
where we have used $ \int_{-\infty}^{+\infty} 
\exp \left[ i \frac{1}{2!}(k_y-k_0)^2 \chi''(k_0)  t \right] dk_y=
\sqrt{\frac{2\pi}{t\chi^{\prime \prime}(k_0)}}$.
The solution can be written as
\be
\label{final}
\psi(x,y,t) = \frac{\Psi_0 \sigma_y \sqrt{\pi}}{4\pi}
\cos(ax) F(k_0) \exp \left[i\chi(k_0)t\right]
\sqrt{\frac{2\pi}{t\chi^{\prime \prime}(k_0)}}
\ee
\refeqb{final} can be used to predict the behaviour of wave
amplitude at a given location in space. The various quantities in this
equation are as follows;
\[ k_0 =  \left \{\frac{y^2}{8B_0^2 t^2}
  + \frac{y}{2B_0 t} \sqrt{ \frac{y^2}{16 B_0^2 t^2}
+ \frac{a^2}{2}} -\frac{a^2}{2} \right\}^{1/2} \]
\[ F(k_0) = \exp \left(-\frac{\sigma_y^2 k_0^2}{4} \right) \]
\[ \chi(k_0) = B_0k_0(k_0^2+a^2)^{1/2} +k_0 \frac{y}{t} \]
\[ \chi''(k_0) = B_0 \left[ \frac{3k_0}{\sqrt{k_0^2+a^2}}-
 \frac{k_0^3}{(k_0^2+a^2)^{3/2}}\right] \] 

We next numerically integrate \eqns{psi}{b} with the help of
pseudospectral code to determine the amplitude of the propagating
whistler wave in the linear regime. This amplitude will then be
compared with the one predicted by \refeqm{final}.  The numerical code
is based on the discrete Fourier representation in $k$-space in the
two directions viz, $x$ and $y$. The boundary conditions in the two
directions are periodic in nature. The linear part of the equations is
exactly integrated in Fourier space. On the other hand, the evaluation
of nonlinear part is carried out in real space and then Fourier
transform it to $k$-space. We use Fast Fourier Transform (FFT)
routines to go back and forth in the real and $k$-space at each time
integration step.  The time advancement is done with the help of a
second order leap-frog predictor corrector scheme. The numerical
accuracy of the results is checked continuously by monitoring the
energy conservation laws of EMHD equations at each time interval.  We
evolve $64^2$ and/or $128^2$ Fourier modes in the two dimensional box
of size $20\pi\times20\pi$ and measure the amplitude of wave at a
spatial location determined by the computational box co-ordinates
($x=9.5\pi,y=9.5\pi$).

 \begin{figure}
\begin{center}
\epsfig{file=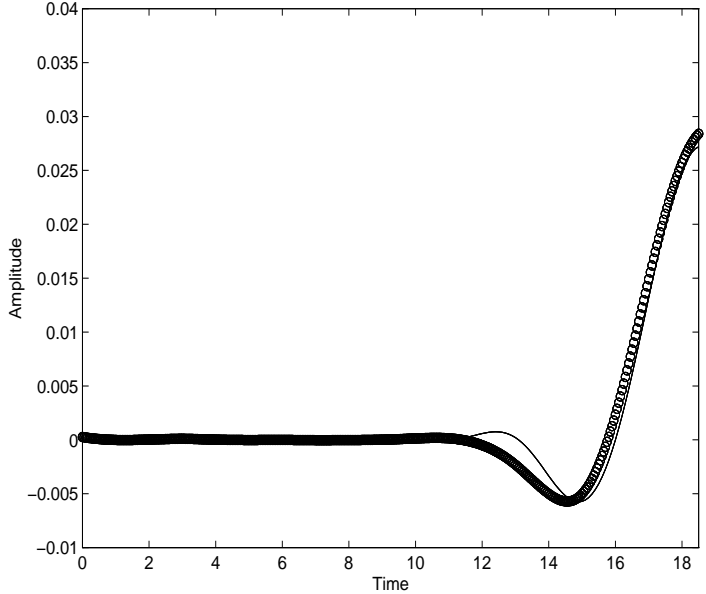, width=10.cm, height=8.cm}
\end{center}
    \caption{The behaviour of  linear wave amplitude. Circle shows analytical
prediction of wave amplitude, which overlaps on the amplitude obtained
from numerical simulation of \eqns{psi}{b} in the linear regime. Clearly,
the analytic amplitude and that obtained from numerical simulation show
an excellent agreement in a long time evolution.}
    \label{c2fig1}
  \end{figure}

A comparison of the amplitude obtained from linear simulation and
stationary phase method is shown in \Fig{c2fig1}. It can be seen from
the figure that the two results are in {\em quantitative} agreement
with each other.  Such a close agreement between analytical and
numerical results thus clearly demonstrates the validity of the our
simulation code in a linear regime.

\section{Propagation Theory of Whistler Wave - Exact Wave Solutions}
We now investigate the case of a propagating whistler wave packet,
which is excited by a current-carrying linear wire antenna. Such a
study is motivated primarily by the work described in [1-10] and
specifically by the experimental observation of Ref. [\cite{sten5}].
In these experiments [\cite{sten5}], it has been observed that the
linear whistler waves can be excited in EMHD plasma, when a
current-carrying linear antenna is kept across an external magnetic
field. The propagating wave packets assume conical shape whose
parallel group velocity is more than the perpendicular
one. Furthermore, the wave propagation has not been observed in their
experiment, when the exciter was placed along the direction of an
external magnetic field.  In order to understand the propagation
characteristics of whistler waves, we carry out both analytical as
well as numerical calculations in the linear regime.  We first study
the evolution of wave pattern using analytical treatment.

The initial conditions are as follows;
\be 
\label{init}
\left. \begin{array}{ll}
\psi(x,y,0) = \Psi_0\exp\left(-\frac{x^2+y^2}{\sigma^2}\right)  \\[.7cm]
b(x,y,0) = 0 \end{array}\right\} 
\ee
where $\Psi_0$ and $\sigma$ are amplitude and width of the
distribution of the initial wave packet respectively. \refeqb{init}
represents constant circular ($\psi$) contours of wave magnetic field
in the plane of variation (i.e., in the $xy$-plane).  Such
perturbations in the $xy$-plane can arise as a result of a
current-carrying wire antenna, which is kept parallel to the
$\hat{z}$-direction. Hence there is no perturbed magnetic field along
this direction. The initial condition for the axial component of the
wave magnetic field is chosen to be zero, $b(x,y,t=0)=0$.
\subsection{Analytical Treatment :}
We Fourier transform \refeqm{init} using the definition of
\refeqm{ft} and  substitute it in \refeqm{psik} to get,
\[ \psi_k(t) = \frac{\Psi_0\sigma^2}{4} \exp\left(-\frac{\sigma^2k^2}{4}\right)
\left\{ \exp(\lambda_+t) +\exp(\lambda_-t) \right\} \]
We inverse Fourier transform this equation and solve in the limit of
$kd_e\ll1$ and $\mu=0$.  Under these limits, the eigen values are
$\lambda_{\pm} = \pm iB_0 k_y k$, and the above eqn becomes
$\psi(x,y,t) = I_++I_-$, where
\[ I_{\pm} = \frac{\Psi_0\sigma^2}{8\pi}
\int _{-\infty}^{+\infty} \int_{-\infty}^{+\infty}dk_xdk_y
\exp(ik_xx+ik_yy) \exp\left(-\frac{\sigma^2k^2}{4}\right)
\exp(\pm iB_0 k_y k t) \]
We first solve $I_+$. Using the transformations $k_x = k \cos\theta$
and $k_y = k \sin\theta$, which give $dk_x dk_y = k ~dk ~d\theta$. The
angle $\theta$ varies from $0$ to $2\pi$, while $k$ varies from $0$ to
$\infty$. The integral $I_+$ then takes the form,
\eqa
 I_+ = \frac{\Psi_0\sigma^2}{8\pi}
\int _{0}^{+\infty}\int_{0}^{2\pi}kdkd\theta
 ~\exp\left(-\frac{\sigma^2k^2}{4}\right)
\exp\left[i(ky+B_0 k^2t) \sin\theta \right]\\
 \times \exp\left(ikx\cos \theta\right)
\eqe
The exponential term containing sine in the above expression can be
expanded using Bessel identity \cite{grad}, $\exp(i\xi\sin\theta) =
\sum_{n=-\infty}^{+\infty} J_n(\xi)\exp(in\theta)$.  We thus get
\eqa
I_+ = \frac{\Psi_0\sigma^2}{8\pi}
\int _{0}^{+\infty}\!\int_{0}^{2\pi}kdkd\theta
 \exp\left(-\frac{\sigma^2k^2}{4}\right)
\sum_{n=-\infty}^{+\infty} J_n(ky+B_0 k^2t )\\
\times \exp\left[i(n\theta+kx\cos \theta)\right]
\eqe

\begin{figure}[t]
\begin{center}
\epsfig{file=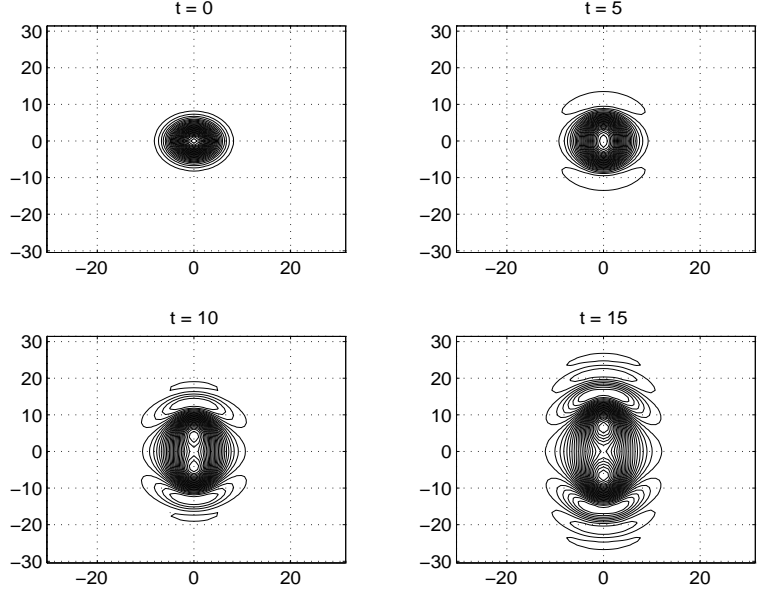, width=10.cm, height=8.cm}
\end{center}
\caption{Propagation of whistler waves. The wave fronts have conical
shape. Figure shows constant $\psi$-contours at various time.  The
vertical axis represents $y$-axis, along which $B_0$ is
oriented. Horizontal axis indicates $x$-axis. Similar wave field
topology has been reported in Ref [\cite{sten5}].}
\label{c2fig2}
\end{figure}   

\begin{figure}[t]
\begin{center}
\epsfig{file=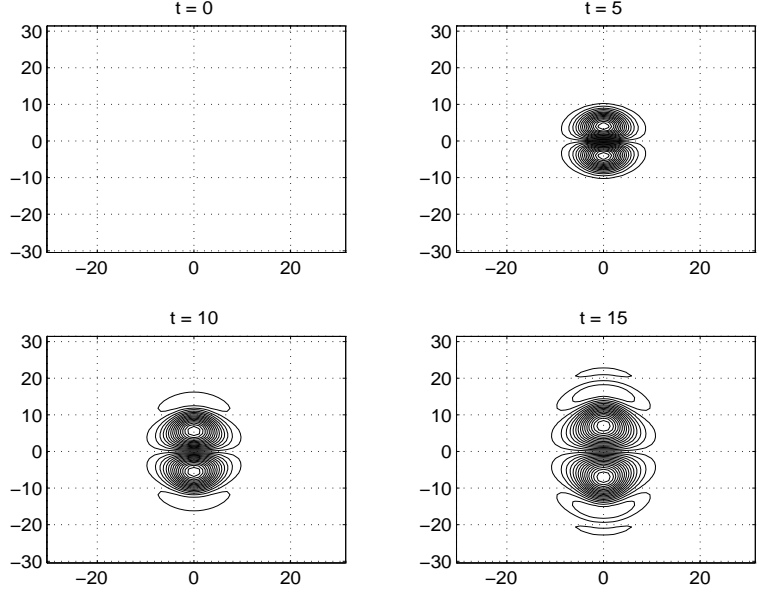, width=10.cm, height=8.cm}
\end{center}
\caption{Constant $b$-contours of magnetic field of propagating whistler waves.
Note that the $b$ component is zero initially, and is generated later by means
of linear mode coupling process.}
\label{c2fig3}
\end{figure}

We then carry out the $\theta$-integral with the help of the identity
[\cite{grad}], $\frac{i^{-n}}{2\pi}\int_{0}^{2\pi}\exp[i(n\theta+\beta\cos\theta)]
d\theta = J_n(\beta)$, and get the equation,
\[ I_+ = \frac{\Psi_0\sigma^2}{4}
\int _{0}^{+\infty}kdk
 \exp\left(-\frac{\sigma^2k^2}{4}\right)
\sum_{n=-\infty}^{+\infty} J_n(ky+B_0 k^2t )
\exp\left(\frac{in\pi}{2}\right)J_n(kx) \]
Using one of the Bessel identities [\cite{grad}], we then evaluate the summation of
above equation.  This gives us
\[ I_+ = \frac{\Psi_0\sigma^2}{4}
\int _{0}^{+\infty}kdk \exp\left(-\frac{\sigma^2k^2}{4}\right)
J_0\left( \sqrt{(xk)^2+(yk+B_0k^2t)^2 } \right) \] We carry out the
similar analysis to integrate $I_-$ and get the total $\psi$ as
follows,
\be
\label{ana_psi}
\psi(x,y,t)= \frac{\Psi_0\sigma^2}{4}
\int _{0}^{+\infty}kdk \exp\left(-\frac{\sigma^2k^2}{4}\right)
\left[ J_0(\Pi_+)+J_0(\Pi_-)  \right]
\ee
Similarly, we solve for \refeqm{bk} and get the expression for the
 axial component as follows,
\be
\label{ana_b}
b(x,y,t)= \frac{\Psi_0\sigma^2}{4}
\int _{0}^{+\infty}dk \exp\left(-\frac{\sigma^2k^2}{4}\right)
\left[ J_0(\Pi_+)-J_0(\Pi_-)  \right]
\ee
where $\Pi_{\pm} = \sqrt{(xk)^2+(yk \pm B_0k^2t)^2 }$.
\beqns{ana_psi}{ana_b} form an exact whistler wave solution of the EMHD model
representing respectively the evolution of poloidal and axial
components of wave magnetic field. The substitution of
$t=0$ in \eqns{ana_psi}{ana_b} retrieves the initial condition i.e.,
\refeqm{init}. The numerical integration of \eqns{ana_psi}{ana_b}
at various time are displayed in \figs{c2fig2}{c2fig3}. It is clear
from the figures that the propagating wave fronts are conical in
shape.  It can be noticed from \Fig{c2fig3} that the axial component,
though absent initially (i.e., at $t=0$), acquires a finite amplitude
during propagation.

The axial component of wave field is generated primarily due to linear
coupling in the equations representing the evolution of $b$ and $\psi$
(i.e., \eqns{b-eqn}{psi-eqn}).  The topology of the magnetic field in
the propagating wave packet is consistent with Ref [\cite{sten5}].

\section{Simulation of Linear Wave Propagation}
We now carry out numerical simulation of low amplitude whistler waves.
The initial condition is the same as that used earlier in the
analytical treatment.  However, unlike the linearized approximation
(\eqns{b-eqn}{psi-eqn}) used earlier to obtain the analytical
expression for the evolved fields, here we make use of the exact
\eqns{psi}{b} for simulation. At small amplitude we expect the linear
approximation to be valid, and hence an agreement of the simulation
results with those presented in the earlier section.

\begin{figure}[ht]
\begin{center}
\epsfig{file=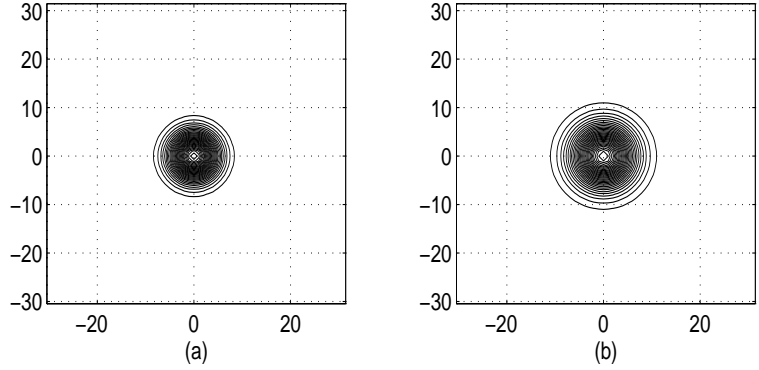, width=10.cm, height=5.cm}
\end{center}
\caption{Diffusion of wave magnetic field, when the exciter is parallel 
to $B_0$.  (a) represents constant contours of perturbed magnetic
field at $t=0$ and (b) shows the diffusion of initial field at $t=20$
unit of time.}
\label{c2fig4}
\end{figure}  
  
In the absence of $B_0$, the fields simply diffuse as shown in
\Fig{c2fig4}. This configuration can be viewed as identical to that
of the Ref [\cite{sten5}] wherein the linear wire antenna is kept
parallel to the external magnetic field. The field lines merely
diffuse in such a scenario. The other case of linear wire antenna
being placed across an external magnetic field can be simulated by
choosing $B_0$ as finite. In this case we observe propagating whistler
wave packets with asymmetric orientation in the $xy$-plane. The phase
fronts have conical shape as reported both in our approximate linear
analytical treatment of the problem as also in experiments. This can
be understood by realising that the fronts move faster along $\hat{y}$
(direction of external magnetic field) than across. This also shows
clearly that the group velocity along $B_0$ is larger than the
perpendicular group velocity.
\begin{figure}[h]
\begin{center}
\epsfig{file=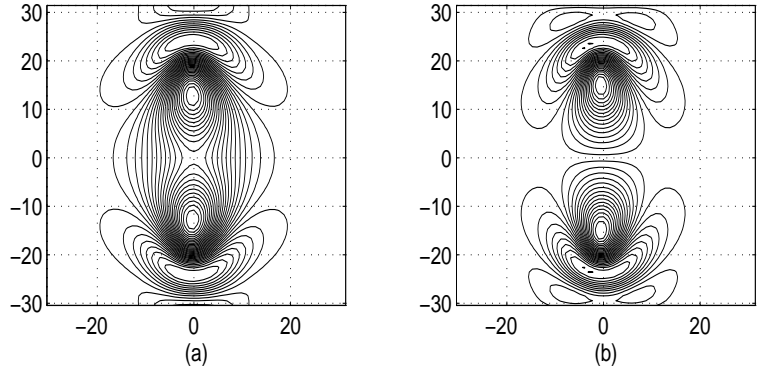, width=10.cm, height=5.cm}
\end{center}
\caption{Propagation of whistler waves recorded at time $t=20$.
 (a) $\psi$-field, (b) $b$-field contours.}
\label{c2fig5}
\end{figure}    
A quantitative estimation of the group velocities of the propagating
whistler wave packet in the two directions i.e., parallel and
perpendicular to $B_0$ can be carried out as follows. The two group
velocities can be computed from the dispersion relation.
The perpendicular group velocity ($V_{G_\perp}$) can
be given as
\be
\label{vgx}
V_{G_\perp} = \frac{\partial\omega}{\partial k_x} =\frac{B_0 k_y k_x}{k
(1+k^2)} -
\frac{2 k k_x k_y B_0}{(1+k^2)^2}.
\ee            
The typical wave numbers $k_x$ and $k_y$ in the above equation can be
estimated as $\sigma_x^{-1}$ and $\sigma_y^{-1}$ respectively, which
represent the width of initial Gaussian distribution of the wave
packet. Replacing $k_x$ and $k_y$ by $\sigma_x^{-1}$ and
$\sigma_y^{-1}$ respectively in \refeqm{vgx}, we get
\be
\label{vx}
 V_{G_\perp} = \frac{B_0/\sigma_x \sigma_y}{\sqrt{\frac{1}{\sigma_x^2}+\frac{1}{\sigma_y^2}}
\left\{ 1 + \left(\frac{1}{\sigma_x^2}+\frac{1}{\sigma_y^2} \right) \right\}} 
-\frac{\frac{2B_0}{\sigma_x
\sigma_y}\sqrt{\frac{1}{\sigma_x^2}+\frac{1}{\sigma_y^2}}} {\left\{ 1
+ \left(\frac{1}{\sigma_x^2}+\frac{1}{\sigma_y^2}\right)
\right\}^2}
\ee
Similarly, the parallel group velocity gives,
\eqnu
\label{vgy}
 V_{G_\parallel} = 
\frac{B_0}{\sqrt{\frac{1}{\sigma_x^2}+\frac{1}{\sigma_y^2}} 
\left(1+\frac{1}{\sigma_x^2}+\frac{1}{\sigma_y^2}\right)^2 }
\left\{ \left(1-\frac{1}{\sigma_x^2}-\frac{1}{\sigma_y^2}\right)\frac{1}{\sigma_y^2}+ \right. \nonumber\\
\left(1+\frac{1}{\sigma_x^2}+\frac{1}{\sigma_y^2}\right) 
\left. \left(\frac{1}{\sigma_x^2}+\frac{1}{\sigma_y^2}\right) \right\}
\eqne
We had chosen in the simulation, $\sigma_x^2=\sigma_y^2=20.0$ and $
B_0=1.0$, which yields $V_{G_\perp}= 0.14$ and $V_{G_\parallel}=0.41$,
thereby clearly indicating that $V_{G_\parallel} > V_{G_\perp}$. It is 
because of the larger parallel group speed, the wave magnetic field
structures are conical and they propagate along the direction of the ambient
magnetic field.

\begin{figure}[h]
\begin{center}
\epsfig{file=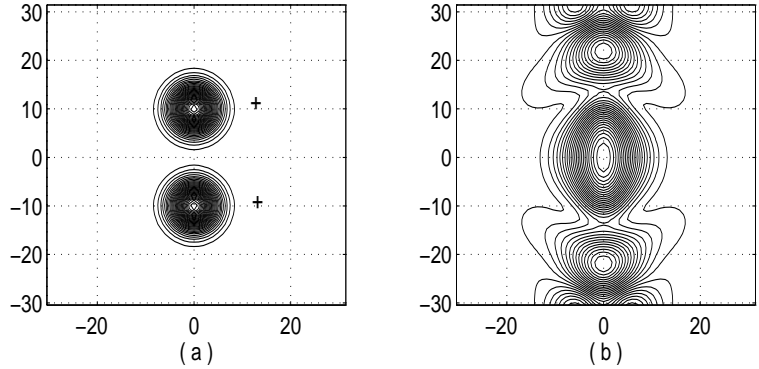, width=10.cm, height=5.cm}
\end{center}
\caption{Interaction of two whistler wave packets excited by parallel
currents in the two antennae show an attraction.  Figure (a) shows the
wave fields at $t=0$ and figure (b) represent the field at $t=20$.}
\label{c2fig6}
\end{figure} 

\section{Collisional Interaction of Two Whistlers}
\begin{figure}[h]
\begin{center}
\epsfig{file=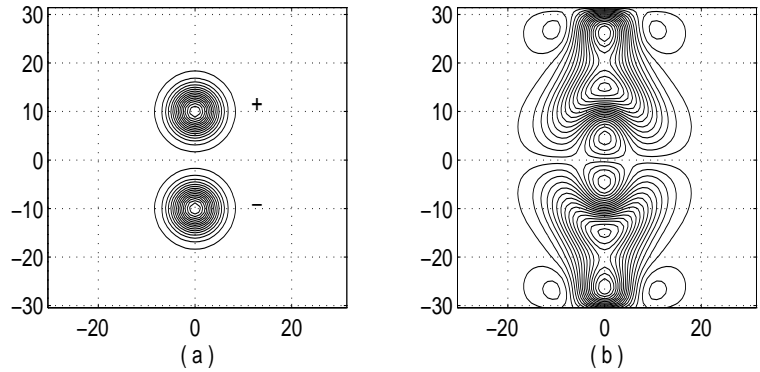, width=10.cm, height=5.cm}
\end{center}
\caption{Two whistler wave packets launched by anti-parallel currents
in the two wire antennae repel each other. Figures (a,b) show the wave fields
respectively at $t=0$ and $t=20$.}
\label{c2fig7}
\end{figure}

We now study the collisional interaction between whistler wave
packets.  Such interactions can possibly exist in the Earth's
ionosphere where the oppositely traveling waves along the Earth's
dipole magnetic field collide. In a laboratory plasma, such an
interaction can be studied by exciting the wave field from the two
linear wire antennae separated by a finite distance.  The currents
carried by the oppositely traveling waves are either parallel or
anti-parallel to each other. Our objective here is to investigate as
what kind of resultant state would be borne out from such
interactions.  When initial excitation is due to the antennae current
flowing in the same direction (i.e., parallel currents in the
antennae), they give rise to wave fields of similar signs in the
plasma. This is shown in \Fig{c2fig6}(a). The evolution of the two
wave packets then leads to the formation of a magnetic null line. The
magnetic pressure being weak at the central region, there is a
tendency of attraction between the magnetic configuration caused by
the two currents. Along with this attraction as the whistler waves get
excited and propagate, the null line at the center alters and forms an
{\it O}-point (see \Fig{c2fig6}(b)). The attraction of wave magnetic
fields can also be understood as follows. The wave magnetic field
induces parallel currents in the plasma, which then attract each
other. This can be contrasted with \Fig{c2fig7} where the initial
excitation are due to the anti-parallel currents in the two wire
antennae. It is observed that there is no such magnetic null line for
this case, albeit here the magnetic field maximizes in the centeral
region, causing repulsion. It can be clearly seen that in this case
there is no formation of {\it O}-point at the center. This is shown
in \Fig{c2fig7}(a,b). In this case, the wave fields induces
anti-parallel currents, which repel each other.

 \section{Nonlinear Whistler Waves} 
\begin{figure}[ht]
\begin{center}
\epsfig{file=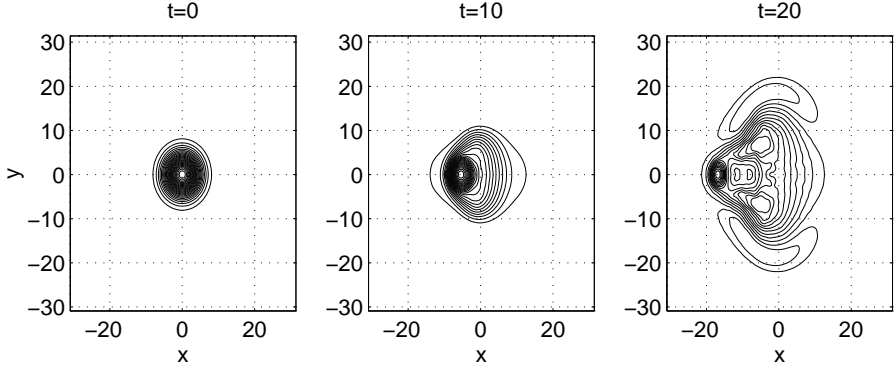, width=12.cm, height=5.cm}
\end{center}
\caption{Large amplitude whistler wave. The wave field is distorted in 
$-x$ direction due to experiencing a net ${\bf J} \times {\bf B}$ force.}
\label{c2fig8}
\end{figure}

In the previous section we have
presented detailed studies on the propagation and collision of
whistler waves, when the amplitude of initial perturbation was smaller
than the external magnetic field ($B_0$). We now carry out our
investigation, when the amplitude of the perturbed field is comparable
or greater than the ambient magnetic field ($B_0$). This is the regime
where nonlinear interactions may be of great importance. Nonlinear
excitation of whistler in laboratory plasmas has n't been yet
understood fully [\cite{sten6}]. We therefore undertake this study to
understand nonlinear evolution of whistlers. Our choice of initial
field perturbation is identical to the one discussed in the previous
sections, which was also consistent with the excitation mechanism
adopted in the experiments [\cite{sten5,sten6}]. In this section we
choose to study the evolution of the similar initial perturbation,
however, now the perturbed field has an amplitude which is comparable
or larger than the ambient field $B_0$. \Figb{c2fig8} shows the
initial and the evolved field configurations. Clearly the magnetic
field topology is distorted and distinct from the linear results
presented earlier (see \Fig{c2fig2}, \Fig{c2fig5}(a)).

The distortion (or modification) in the propagation characteristics of
whistler waves can be understood as follows.  The linear wire antenna,
which induces the wave magnetic field in the $xy$-plane of variation,
essentially carries current (${\bf J}$) along positive
$\hat{z}$-direction i.e., out of plane of the page. A constant
magnetic field ($B_0$) is already present in the
$\hat{y}$-direction. This ambient magnetic field interacts with the
antenna current, and gives rise to a net ${\bf J} \times {\bf B}_0$
force, which is directed along $-\hat{x}$ direction. Since the
nonlinearities are at work in the large amplitude simulation, they
exert the resultant Lorentz force (i.e., ${\bf J} \times {\bf B}$
force) on the wave field. The induced magnetic field of the wave
thereby get influenced by the Lorentz force and consequently show a
tendency to move in the direction of this force. Furthermore, a
combined action of the wave motion along $\hat{y}$-direction and the
push in $-\hat{x}$ direction, overall results in slight distortion of
the wave field topology.  This is indeed what we observe in the
simulation also (see
\Fig{c2fig8} at $t=20$). Our numerical investigations, thus demonstrate that
nonlinear interactions significantly influence the propagation
properties of whistler waves in EMHD.

Nonlinear evolution of whistler waves is widely observed in many
ionospheric and magnetospheric phenomena. For instance, whisther mode
has been reported in the context of equatorial region of Earth's
magnetosphere based on a modulational instability analysis
[\cite{shukla1}], where it was shown that amplitude modulation in
whistlers develops on much faster time scale than the wave transit
period. Nonlinear interaction of whistlers is also likely to occur
with magnetosonic fluctuations [\cite{shukla2}]. On the other hand,
the nonlinear whistlers can potentially excite electrostatic
low-frequency collisional gradient drift modes that are relevant to
understanding the density irregularities in the lower part of the
ionosphere [\cite{shukla3}].  Additionally, electromagnetic whistler
wave, propagating parallel to an external magnetic field, are also
reported to decay into another circularly polarized electromagnetic
wave and an ion acoustic wave in a homogeneous plasma
[\cite{shukla4}].  This result remains unaltered under the combined
effects of the relativistic-mass and ponderomotive-force
nonlinearities [\cite{shukla5}].  Whistler waves in inhomogeneous
media play a critical role in ionospheric density striations
[\cite{shukla6}]. Our simulations of nonlinear evolution of whistler
waves, relevant to understanding a number of features described in the
above work, thus reveal that it is the hall force corresponding to the
${\bf J} \times {\bf B}$ term in electron momentum equation that plays
a crucial role and it is responsible for governing the nonlinear
interactions associated with the electron whistler modes.

\section{Conclusion}
In conclusion, we have carried out a detailed study of whistler wave
propagation with the prime objective to understand their interaction
and the nonlinear propagation characteristics. For this purpose a
variety of initial configuration were envisaged and their evolution
studied. Whistler waves being dispersive, in the linear small
amplitude limit different modes of the wave packet propagate with
different phase velocities. However, the entire wave front moves with
the typical group velocity. The shape of the wave front is finally
governed by the group speed. The linear propagation characteristics
are found to be in agreement with the experimental observations. In
the nonlinear regime we show that the propagation is significantly
altered. This indicates the importance as well as the subtle role of
nonlinearity in the context of whistler wave propagation thereby
countering earlier speculations and statements about robustness of
these modes against nonlinear effects. Our studies presented in this
paper can be useful in understanding a number of processes associated
with the linear whistler waves, their mutual interaction and nonlinear
feature in the context of interplanetary ionospheres and radiation
belts.

\label{lastpage}
\end{document}